\documentclass[prl,twocolumn,superscriptaddress]{revtex4-1}
\usepackage{graphicx}
\usepackage{gensymb}
\usepackage{amsmath}
\usepackage{amssymb}
\usepackage{mathrsfs}
\usepackage{xcolor}
\usepackage{multirow}
\usepackage[normalem]{ulem}
\usepackage{cancel}
\usepackage{siunitx}
\usepackage{graphicx}
\usepackage{dcolumn}
\usepackage{bm}

\newcommand{\farutinremove}[1]{{}}
\begin{document}


\title{
When Blood Parts Ways: Phase Separation in Microstructured Environments}


\author{Sampad Laha}
\affiliation{Department of Mechanical Engineering, Indian Institute of Technology, Kharagpur, India                                     }
\affiliation{Univ. Grenoble Alpes, CNRS, LIPhy, F-38000 Grenoble, France
}
\author{Ananta Kumar Nayak}
\affiliation{Univ. Grenoble Alpes, CNRS, LIPhy, F-38000 Grenoble, France
}
\author{Alexander Farutin}
\affiliation{Univ. Grenoble Alpes, CNRS, LIPhy, F-38000 Grenoble, France
}
\author{Suman Chakraborty}
\affiliation{Department of Mechanical Engineering, Indian Institute of Technology, Kharagpur, India                         }
\author{Chaouqi Misbah}
\affiliation{Univ. Grenoble Alpes, CNRS, LIPhy, F-38000 Grenoble, France
}

\date{\today}

\begin{abstract}
Understanding how red blood cell (RBC) suspensions navigate porous materials is critical for for both fundamental physiology, such as maternal-fetal exchange in the placenta, and transformative biomedical applications, including rapid, low-cost disease diagnostics from a single drop of blood in resource-constrained settings.
Here we elucidate how RBC movement through fibrous microporous structures is influenced by cell aggregation agents, emphasizing the impact of their clustering, membrane flexibility, and confinement. By varying the volume fraction of the RBC (hematocrit) and aggregation strength, we reveal a surprising phase separation: a dense RBC core surrounded by a cell-free layer, an effect not previously reported in whole blood studies. This separation is shown to be more pronounced with rigidified cells and persists even at high hematocrit levels, unlike in healthy samples. By connecting RBC deformability and aggregability to pore-mediated phase dynamics, our study provides a foundation for new diagnostic tools capable of classifying blood disorders or evaluating blood quality using only a sheet of structured paper, seamlessly integrating fundamental fluid mechanics with translational biomedical innovation in a previously unexplored manner.\end{abstract}

\maketitle

Phase separation of blood\cite{clavica2016red}plays a critical role in physiological porous environments, such as the human placenta\cite{chappell2023review,zhou2022red,shimizu2017impact}, where it regulates the distribution of red blood cells (RBCs) and plasma. This, in turn, critically influences oxygen delivery, nutrient transport, and immune function within tissues\cite{lucker2017relative,minton2021red,lucker2015dynamic}.
Equally important is the  role of phase separation in diagnostic paper strips for rapid disease detection in resource-limited settings\cite{morbioli2025clinical,lu2018high,guo2020synthetic}. The functionality of these porous systems is in general dictated by complex local variations in RBC volume fraction (hematocrit) and the flow resistance within inherently disordered porous fluidic pathways, factors that remain poorly understood\cite{zhou2022red,secomb2017blood}. Advancing our understanding of this interplay can not only inform the design of next-generation blood diagnostic technologies with precise control over RBC dynamics\cite{tomaiuolo2014biomechanical,recktenwald2022erysense,laha2022smartphone,biswas2022instrument}, but also provide insights into unique pathological conditions, such as sickle cell disease\cite{ingram1959abnormal,kato2018sickle}.

The central physical issues of phase separation of blood in complex porous environments involve the understanding the dynamical evolution of the RBCs and plasma as they navigate through irregular, confined pore structures with varying sizes and shapes. Key factors include how the mechanical properties and interactions of RBCs, such as their flexibility and adhesion\cite{lominadze2002involvement,baskurt2013erythrocyte,huisjes2018squeezing,skalak1969deformation,fedosov2014deformation} influence their distribution and separation under flow. Local variations in the hematocrit (HCT) affect the blood viscosity and flow resistance\cite{walburn1976constitutive}, while interactions between plasma, RBCs, and vessel walls, including phenomena like margination, play important roles in driving phase separation\cite{secomb2017blood}. 

A major challenge in decoding the complex navigation of blood samples in fibrous porous pathways lies in linking the microscopic dynamics of RBCs with the macroscopic flow features observed\cite{laha2023cellular}. This raises unresolved questions about how intricate pore geometries and heterogeneous micro-environments jointly modulate phase separation patterns, the quantitative relationships between cell mechanics and separation dynamics, and the impact of pathological changes on these processes. 

In this Letter, we present hitherto-unraveled experimental insights into the dynamics of RBC suspensions spreading through porous filter paper, achieved by precisely modulating the combined consequences of cellular aggregation and deformability. Our findings pinpoint previously unreported transitional regimes, ranging from phase separation between cells and suspending fluid to cohesive single-phase spreading, driven by a complex, non-linear interplay among the HCT levels, aggregation intensity, filter pore size, and cellular deformability. By integrating detailed experimental observations with quantitative simulation insights, we dissect the underlying physical mechanisms that remained largely unexplored. Our results not only advance fundamental insights on blood flow in porous media but also offer promising avenues to link micro-scale spreading phenomena with clinically relevant blood disorders characterized by abnormal aggregation or altered cell mechanics, thereby broadening the potential impact across biophysics, biomedical engineering, and disease pathology.

\paragraph{Experimental set-up  and measurement protocol.}

\begin{figure}
\begin{center}
\includegraphics[width=1\columnwidth]{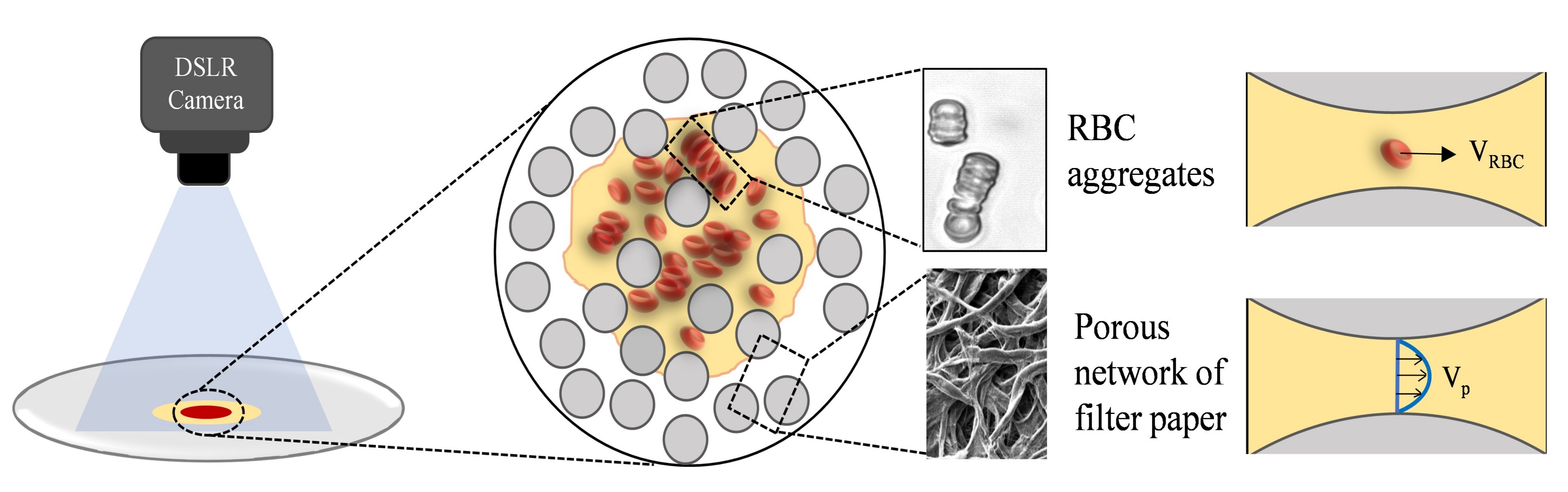}
\caption{
\label{fig1} Schematic representation of the experimental setup consisting of a DSLR camera for video recording of the wicking of RBC suspension on laboratory filter paper. The schematic also shows a representative model of heterogenous porous media with phase separation occurring between the cellular matter and suspending fluid. Magnified images in the inset show optical microscopic images (10X) of RBC aggregates obtained after dextran infusion in PBS, as well as scanning electron microscopic images of the internal porous structure of Whatman grade 1 filter paper (350X). The network of pores forms micro-confinements through which PBS and RBCs flow. }
\end{center}
\end{figure}

Figure \ref{fig1} represents a schematic description of the experimental setup and the model problem investigated. The wicking experiments were conducted on laboratory grade filter papers of different pore sizes (Whatman 1 pore size :11 $\mu$m, Whatman 5 pore size: 2.5 $\mu$m and Whatman 4 pore size: 25 $\mu$m). Dextran 150 KDa was dissolved in phosphate buffered saline (PBS) in appropriate proportions to constitute a stock solution of dextran concentration 50 mg/ml. This stock solution was thereby serially diluted with PBS (Phosphate Buffer Saline) to obtain solutions having different dextran concentrations.
Blood samples of healthy individuals were collected in EDTA (Ethylenediaminetetraacetic acid coated) tubes from ESF (Etablissement Fran\c {c}ais du Sang, Grenoble, France). RBCs were separated from whole blood by standard laboratory centrifugation process and suspended  in dextran-infused PBS solution. 
Experiments were also conducted using RBC suspensions in pure PBS without dextran in order to unveil the key differences in wicking due to dextran-induced aggregation.  In order to study the effect of cell deformability, experiments were also conducted with RBCs rigidified with the help of 0.1$\%$ glutaraldehyde solution in accordance with previously reported protocols \cite{abay2019glutaraldehyde}.
20 $\mu$L of liquid was dispensed on the filter paper using micro-pipettes for each experimental run. The spreading video was captured for a time interval of 2 minutes with a digital camera (Nikon D5600) and thereafter divided into subsequent image frames at an interval of 1 second. Image segmentation is carried out in ImageJ to calculate the phase separation area for different HCT values (see \cite{SI}) . In order to investigate the effect of dextran on the structure of RBC aggregates, we performed microscopic measurements (using an inverted optical microscope in bright field mode) of RBC suspension samples (of HCT 2$\%$) on glass-coverslips covered with a thin layer of Bovine Serum Albumin (BSA) [refer to Fig S1 (a-d) in the Supplementary Material\cite{SI} for more details]. The HCT and dextran concentrations used in this study are reported in Table 1.
All the experimental runs are repeated at least 3 times to ensure statistical reproducibility of the results.
\begin{table}[h!]
  \begin{center}
    \caption{Details about HCT$\%$ and dextran concentrations.}
    \label{tab:table1}
    \begin{tabular}{|c|c|}
     \hline
      Hematocrit  (HCT$\%$) & 15, 18, 25, 30, 35, 40, 50, 55, 60 \\
      \hline
      Dextran  (mg/ml) & 0 (D0), 15 (D15), 30 (D30), 50 (D50) \\
      \hline
    \end{tabular}
  \end{center}
\end{table}

\paragraph{Numerical simulations}
Numerical simulations have been carried out in a two-dimensional porous media with randomly placed circular obstacles by adopting our immersed boundary lattice-Boltzmann (IBLBM) code\cite{Shen17}, in a similar set-up as that of a previous study focusing on modeling blood flow through the porous regions of human placenta\cite{zhou2022red}. A particular obstacle concentration $\psi$,   has been considered; the associated substrate porosity $\phi$  is defined as $\phi= 1-\psi$.  The radius of obstacles is considered { as 3 $\mu$m, close to the radius of an RBC.  }
The viscosity contrast between the inner fluid of RBC and the outer fluid is considered to be 6 (inspired by that of RBCs). The reduced area, defined as $(A/\pi)/(P/2\pi)^2$ (where $A$ is the enclosed area and $P$ is the perimeter) of the RBC is taken to be 0.65. In order to emulate the capillary-driven flow observed in paper-based wicking experiments, which is relatively slower than the externally driven blood flow in microcirculation, we considered a very low flow rate for the external flow (with mean flow velocity of about $0.01$ mm/s). This externally driven fluid flow is characterized by the capillary number with respect to deformation of the RBC membrane, representing the ratio of the viscous strength of the external fluid to the bending strength of the membrane, where bending modulus of the RBC membrane is taken to be $3 \times 10^{-19}$ J. We considered capillary number to be 1 and 0.1 for simulations with healthy and rigidified RBCs respectively. 

\begin{figure}
\begin{center}
\includegraphics[width=0.99\columnwidth]{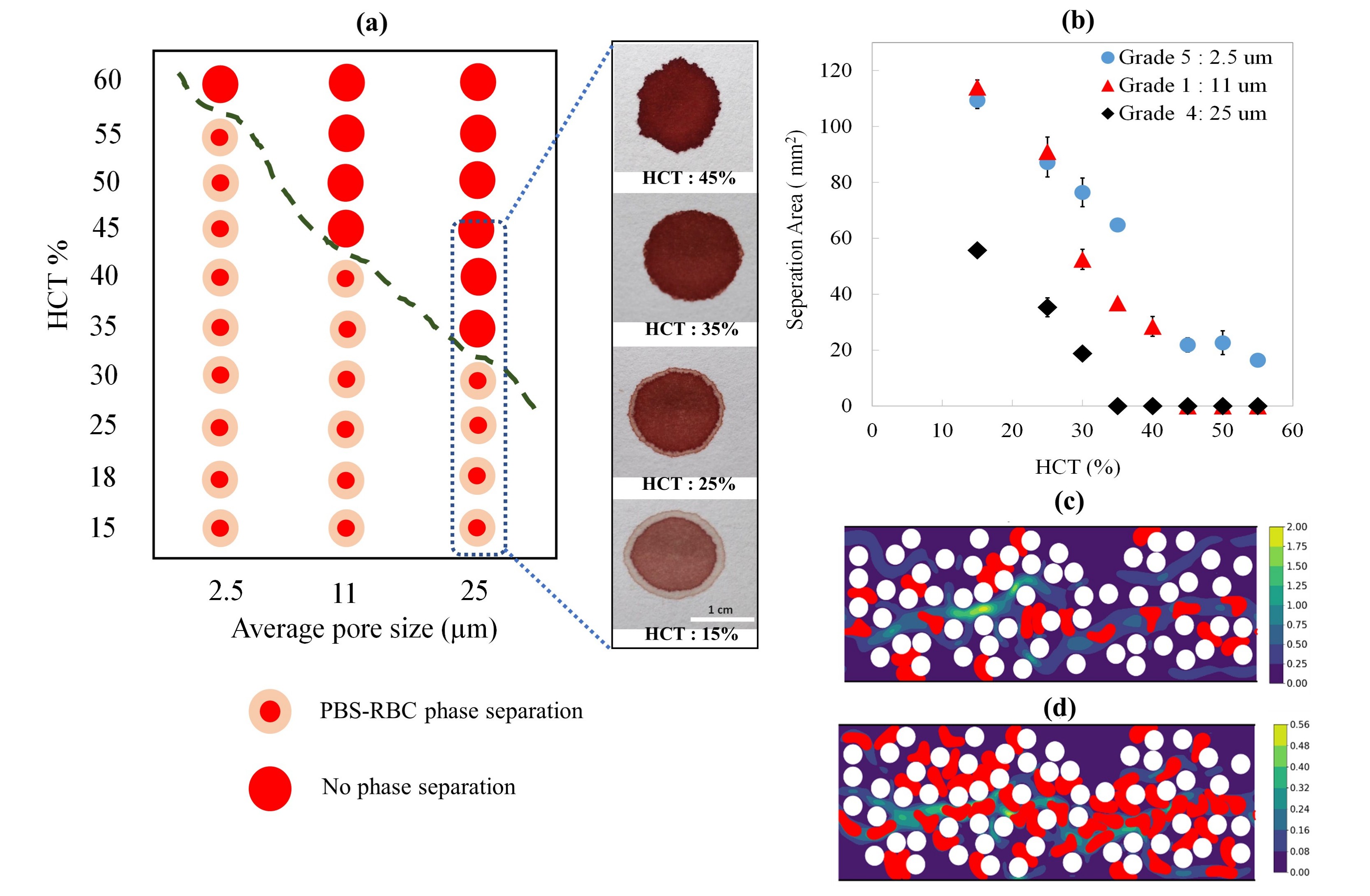}
\caption{
\label{fig2} 
Regime diagrams for RBC-PBS phase separation: (a) shows a regime diagram highlighting the effect of paper pore size on the threshold HCT for phase separation. (b) shows the variation of the phase separation area with sample HCT for 3 different grades of paper with varying pore sizes. (c) and (d) show the 2D porous domain of the numerical results with randomly placed circular obstacles; the figures highlight the difference in RBC distribution during flow of cellular suspensions of HCT 11.8$\%$ and HCT 29.5$\%$ respectively, the medium porosity being ~ 73.9$\%$. As HCT increases, cellular and liquid phase tends to move together without phase separation. The same can be observed in the experimental images (as presented in the inset of (a)) of blood stain patterns on Grade 4 filter paper, showing how phase separation tendency decreases with increase in sample HCT. }
\end{center}
\end{figure}

\paragraph{Emergence of phase separation and the impact of confinement-}  We analyzed spreading of a drop of RBCs suspension on a paper surface. The first noticeable result (Fig.\ref{fig2} a) is that at low enough HCT the spreading of the drop is accompanied by a phase separation: a dense suspension of RBCs in the core region surrounded by a ring of PBS (which is quasi-free of RBC). The PBS ring width decreases by increasing HCT, and vanishes beyond a critical hematocrit. This critical hematocrit depends on pore size: larger pore sizes have the tendency to restrict phase separation to lower HCT. Figure \ref{fig2}b shows the area of the PBS annulus as a function of HCT. 

 PBS, the suspending fluid, is almost rheologically analogous to water and hence endures significantly less viscous resistance as compared to flow of packed cellular matter through narrow constrictions. This may tempt us to believe that  the larger HCT is, the higher will be the degree of separation, since PBS could go between RBCs interstices, even at high concentration, moving thus faster than crowded RBCs. However, from our experimental analysis, we found something quite contrasting.
 
Figure \ref{fig2}b shows that as we increase the HCT of the D0 sample for Grade 1 paper, the separation area tends to decrease up to ~ 40$\%$ HCT, above which no phase separation is observed. As paper is a heterogenous porous medium, there are several areas where the pore size is smaller than the size of an RBC; consequently, RBCs flowing through these regions are trapped. In low HCT samples, most of the RBCs are trapped in these slow-flow constricted regions. The remaining cells flow through the less resistant pathways, where RBCs can deform and pass through the pores. This creates enough space for PBS to flow unimpeded through the paths of least resistance, consequently leading to phase separation. The same has been depicted through the numerical investigations carried out using LBM framework, as shown in the 2D porous domain images in Fig.\ref{fig2}d-e (more quantitative numerical results are presented below) . 
By increasing the HCT, a higher proportion of RBCs are trapped, compared to the lower HCT samples, thereby impeding the overall liquid flow across the domain. The overcrowding  of cells causes the RBCs to migrate into the less resistant PBS pathways, where PBS has to push against RBC crowded domains to move forward, causing RBCs to move together with PBS, suppressing thus phase separation. As the pore size increases, the local capillary number (ratio between imposed stress and local RBC shear elastic stress) is also enhanced. Therefore, RBC are more deformed allowing them to flow faster through the tortuous medium, so that  the difference in speed with PBS is lower as compared to the smaller pore size paper; 
The reduced confinement effect therefore allows the RBCs to migrate into the PBS pathways at a much lower HCT level compared to that for smaller pore size papers, consequently leading to the decrease in HCT threshold for phase separation. 

\begin{figure}
\begin{center}
\includegraphics[width=0.95\columnwidth]{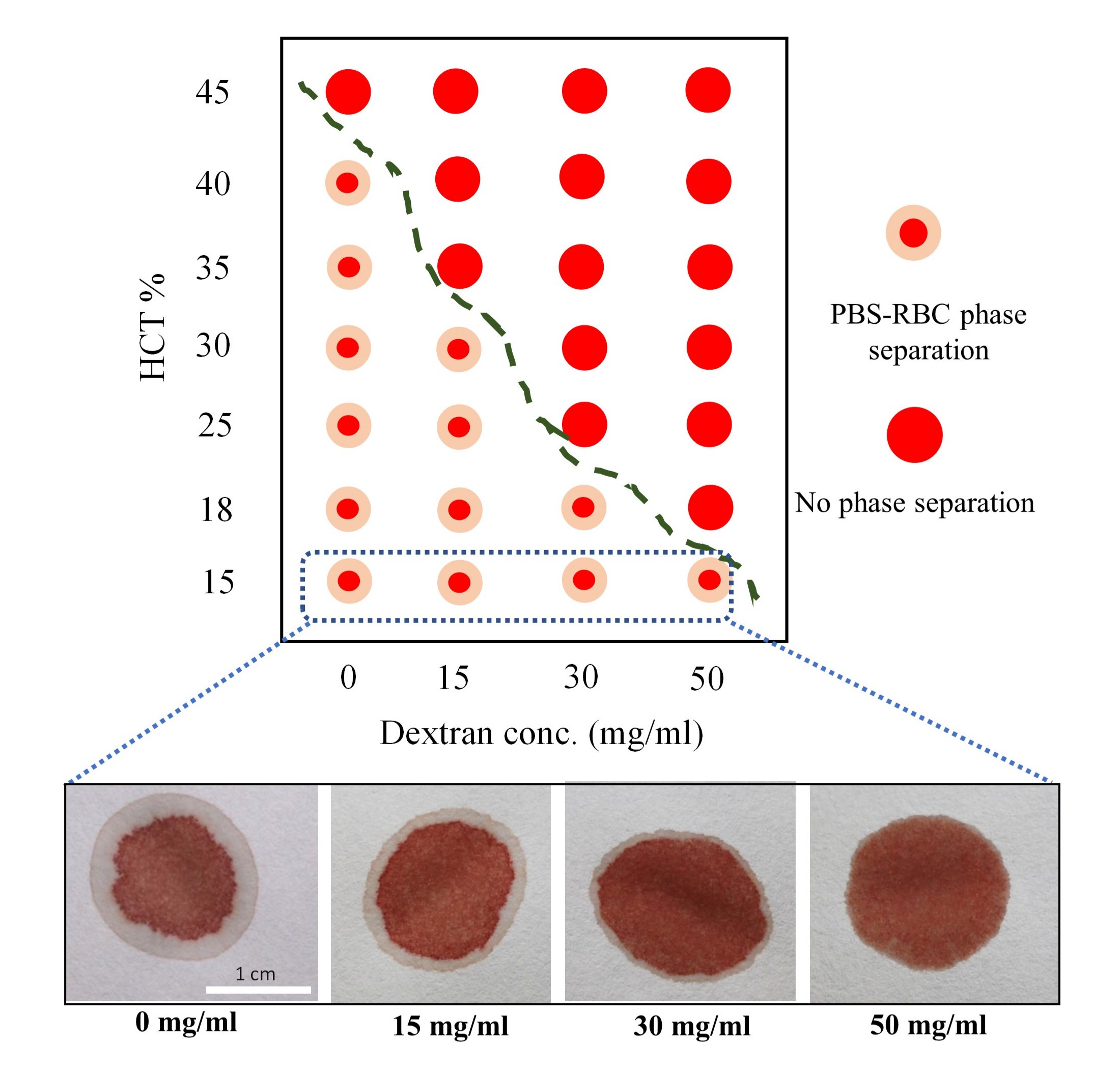}
\caption{
\label{fig3} Regime diagrams for RBC-PBS phase separation: (a) shows the regime plot demonstrating the effect of aggregability on the threshold HCT for phase separation as a function of dectran concentration.
The inset figure shows experimental images of the blood stain pattern formed on Whatman 1 filter paper, clearly demonstrating the decreasing phase separated area with increase in dextran concentration. 
}
\end{center}
\end{figure}

\paragraph{Impact of  RBC aggregation on phase separation}
An outstanding property of blood is the ability of RBCs to adhere to each other, basically due to plasma proteins (e.g. fibrinogen). This adhesion is also present under physiological conditions, albeit reversible. In several diseases (such as diabetes\cite{groeneveld1999relationship,rogers1992decrease}) the aggregation can be so strong so that it becomes irreversible, causing even vessel occlusions. We have analyzed the phase separation problem in the presence of RBC adhesion, caused by dextran molecules\cite{CHIEN1973155} with increasing concentration.
%
 As we increase the dextran concentration, the  RBC aggregates gradually become larger in size as well as more complex in shape (Fig S1(b-d) in \cite{SI}). 
Figure \ref{fig3} shows a regime diagram for phase separation between RBC and PBS, highlighting the role played by dextran concentration on the HCT threshold for phase separation. 

The infusion of dextran within the suspending medium, leads to RBC aggregation\cite{wagner2013aggregation,neu2008effects}. Cellular aggregation causes  more frequent random pore blockages within the porous domain \cite{laha2023cellular} leading to the overall reduction in the diffusive flux. 
The PBS has to push against the increased crowding of RBCs due to aggregation, which causes the suspension to diffuse uniformly without phase separation across most of the HCT values, except for very small volume fractions. These facts explain the decrease of phase separation region upon increasing dextran concentration (Fig.\ref{fig3}).

\begin{figure}
\begin{center}
\includegraphics[width=0.95\columnwidth]{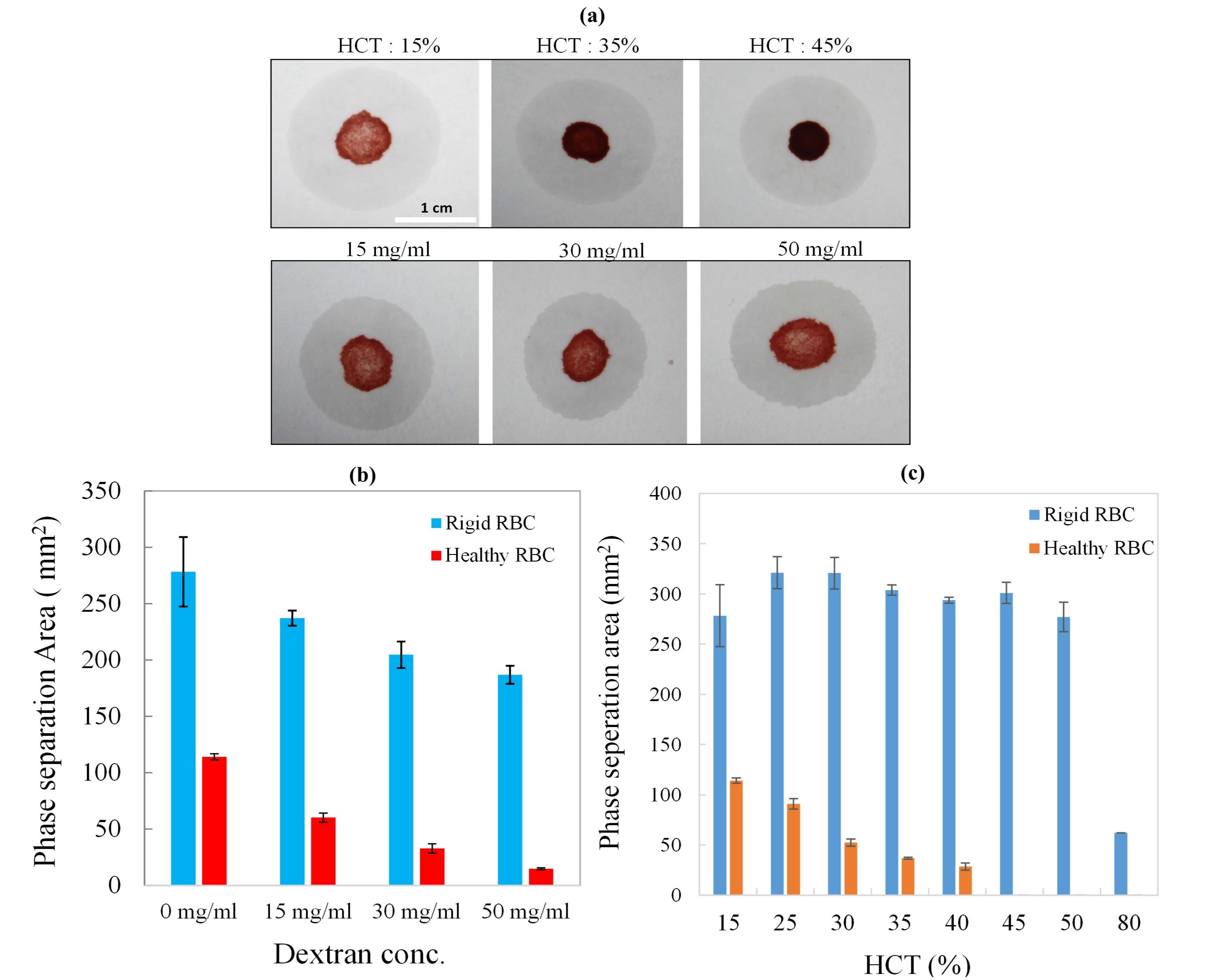}
\caption{
\label{fig4} Effect of deformability of RBC membrane on the phase separation of RBC-PBS suspension. (a) shows the experimental snapshots of blood wicking patterns for different HCT as well as dextran concentrations (with rigid cells), all showing significant phase separation unlike the case of healthy cells. (b) shows variation of phases separation area with HCT$\%$ for zero dextran (D0) case both for rigid and healthy RBC suspension (c) shows the variation of phase separation area with dextran concentration for HCT 15 $\%$, both for rigid and healthy RBC suspension.  }
\end{center}
\end{figure}

\paragraph{Effect of RBC membrane deformability}  
Alteration of cell deformability is a signature in many blood diseases, such as malaria, and diabetes \cite{Tomaiuolo14}. It is thus interesting to see how this factor can impact phase separation. We have studied this question for hardened RBCs with and without dextran. A first remarkable finding is that the phase separation area is much larger for hardened than for healthy cells (Fig.\ref{fig4}a,b), both in the presence and absence of dextran. A second outcome is that phase separation persists always even at extremely high hematocrits (Fig.\ref{fig4}a,b). These findings constitute interesting macroscopic signatures to detect  altered RBCs mechanical properties.

Let us first focus on the understanding of the case without dextran. In this case two factors enhance phase separation: (i)
rigid RBCs tend to tumble near the walls, instead of tank-treading, thereby enduring a greater drag force compared to healthy ones, eventually leading to decreased average speed in microvasculature, compared to deformable cells  \cite{stathoulopoulos2022flows,boas2018assessment}. This effect maintains larger speed differences between PBS and the cells (ii) The lingering effect is usually attributed to cell deformability: when RBC impacts a bifurcation impact, its ample deformation allows it to conform to the obstacle shape causing a certain residence time  of the RBC at the apex\cite{BALOGH20172815,kihm2021lingering}. In contrast, hardened shape do not have such ability to conform and should have much weaker lingering, reducing  cell crowding, which would have otherwise slowed down the suspending liquid\cite{rashidi2023red}.

Moreover, in the presence of dextran, it has been 
 reported that with decrease in cellular deformability, the propagation of point-to-point contact to form surface-to-surface contact between the RBCs is severely affected, eventually decreasing the degree of aggregation between the cells \cite{shiga1990erythrocyte}. This is also validated in this work using microscopy of the rigid RBC suspension with different levels of dextran concentration (see SI). While for healthy RBCs, complex-shaped large aggregates were observed for higher dextran concentrations like 50 mg/ml (Fig S1(d)), in case of rigid RBCs, large aggregates were not formed even for 50 mg/ml dextran concentration (Fig S1(e)); only small linear aggregates could be observed. The more complex aggregates form, the more will be the blockage induced retardation of the liquid phase, eventually leading to attenuation of phase separation. Hence, with the impairment of aggregation phenomenon in case of rigid cells, it is expected that PBS flows unobstructed with higher velocity compared to the cells. 
The combined effect of the increase in speed of PBS and decrease in the speed of RBCs lead to the increase in phase separation tendency for rigid RBCs.

\paragraph{Numerical results --} Here we study the  suspension dynamics by numerical simulation by considering a pressure driven flow in a porous medium. 
In order to express the existence of phase separation, we evaluate the spatial as well as time-averaged speed of RBC as $\bar V_{RBC}= {\int {1\over n} \sum_1^n \lvert V_{RBC} (t) \rvert dt \over \int dt } $ 
and that of the suspending fluid as $\bar V_{p}= {\int [ {\int \lvert V_{p} (x,t) \rvert d\Omega \over \int d\Omega } ] dt \over \int dt }$ .  Here $V_{RBC}(t)$  is the velocity of a single RBC at time $t$, $V_{p} (x,t)$    is the fluid velocity at a point $x$ within the porous domain and $n$ is the total number of  RBCs in the domain $\Omega$.  

\begin{figure}
\begin{center}
\includegraphics[width=0.95\columnwidth]{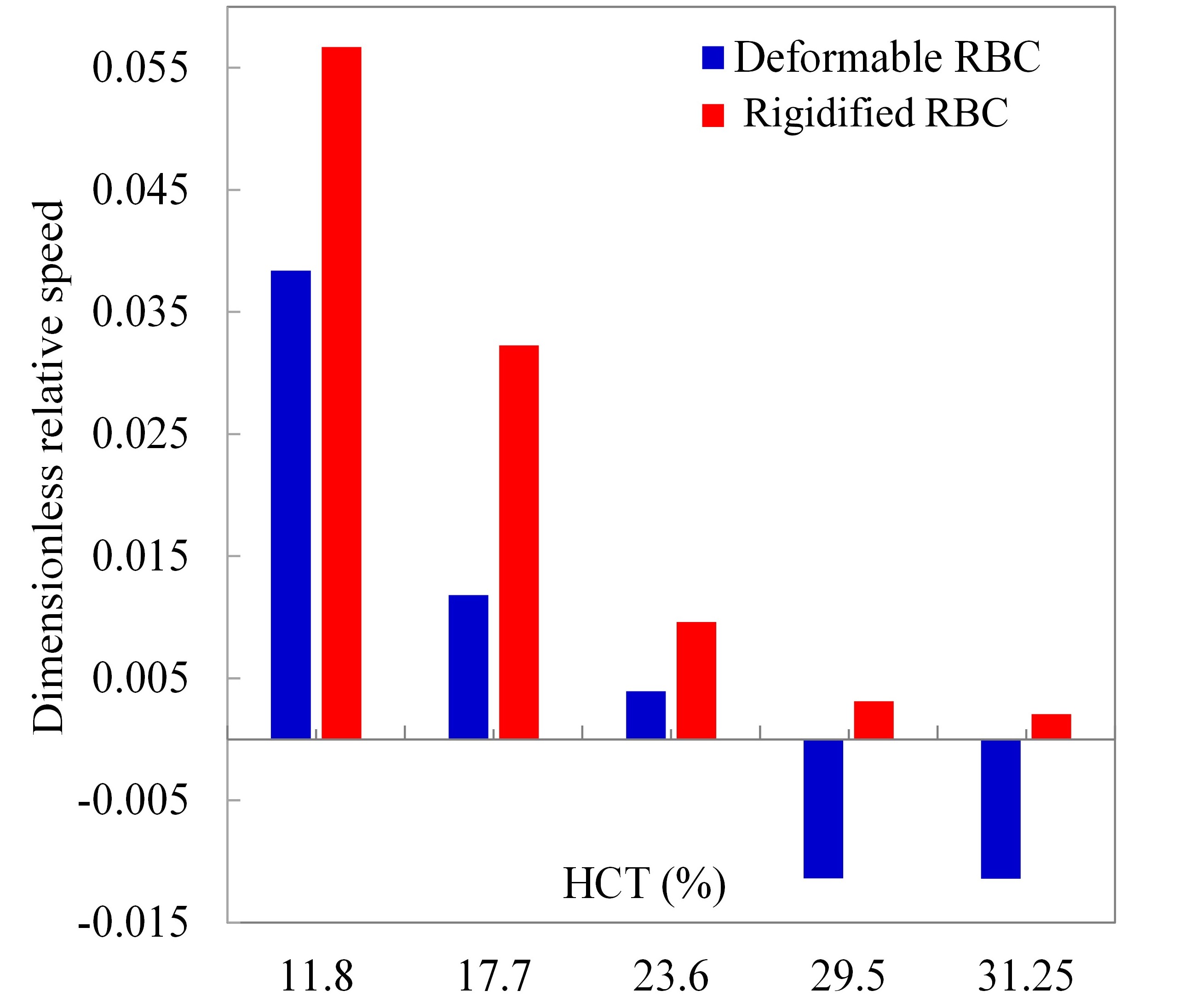}
\caption{
\label{fig5} Numerical results for the relative mean speed between suspending fluid and RBCs for different HCT and for healthy and rigidified cells. For a positive value of the relative speed there is phase separation.  
}
\end{center}
\end{figure}

Phase separation signature is measured by  the dimensionless relative speed between RBC and PBS,$(V_{p}- V_{RBC})/V_{pmax}$, ($ V_{pmax}$ is the maximum imposed inlet velocity of the liquid phase without RBC). A positive value of this quantity means the suspending fluid goes faster than the RBCs, leading to phase separation.
By studying different HCT fractions (as shown in Fig.\ref{fig5}), it is found  that for healthy RBCs, a front cross-over is obtained for $\approx$ HCT 29$\%$ while for rigid RBCs, PBS speed continues to dominate the RBC speed even for higher HCT fractions, exactly similar to what is experimentally observed. 


\paragraph{Discussion and conclusion--}
We report on a previously unrecognized mechanism by which the structural heterogeneity of paper profoundly alters fluid transport dynamics in complex suspensions such as blood. The inherent micro-structural heterogeneity leads to formation of cavities within the porous domain which are filled and subsequently drained out, giving rise to an initial rapid phase of wicking followed by a slower redistribution phase\cite{hertaeg2019dynamics}. Surprisingly, in the absence of phase separation, the advancing front of the fluid was found to remain stationary (i.e. fixed during time) once the droplet is absorbed, a finding that defies the intuitive expectations premised on the capillary action alone.
The observed counterintuitive phenomenon stems from the selective absorption of the suspending fluid into smaller pores, leaving larger cavities partially air-filled. Above a critical HCT, RBCs accumulate in these cavities such that their spacing approaches or falls below the average pore size. This geometry-driven configuration promotes capillary-driven retention of the suspending fluid at the RBC-rich front, effectively suppressing the phase separation and maintaining the fluid-cell contact. Remarkably, we demonstrated that increasing the average pore size significantly alters this balance: while the cavity size increases, so does its capacity to hold more cells, preserving inter-cellular spacing and enabling enhanced fluid retention. This explains the observed decrease in the critical HCT threshold with increasing pore size, a result with no prior analogue in the reported wicking or phase separation literature\cite{hertaeg2020radial,hertaeg2018effect,deprez2019evaluation,frantz2020quantitative,laha2023cellular}.
Our integrated approach, combining precise experiments with the LBM simulations, provides the first mechanistic insight into how microstructural features of a porous medium govern phase separation in cell-laden fluids. This enabled the establishment of a new framework to understand and control phase separation via pore-scale architecture, with direct implications for the design of low-cost, paper-based biomedical diagnostics. Beyond immediate applications, these findings further open new avenues in soft matter physics, transport in disordered media, and microvascular physiology. By linking the RBC deformability and aggregability to pore-mediated phase dynamics, our study lays the groundwork for developing novel diagnostic tools that can classify blood pathologies or assess blood quality using nothing more than a sheet of structured paper, thus bridging interfacial fluid dynamics with translational biomedical innovation in a manner that could not be envisaged thus far.

 All the authors acknowledge financial support from  CEFIPRA (project NO. 6409-3). A. F., and C. M. thank CNES (Centre National d’Etudes Spatiales) and the French-German university program ''Living Fluids'' (grant CFDA-Q1-14) for a financial support. S.C. acknowledges the Department of Science and Technology, Government of India, for J.C. Bose National Fellowship. S. L. acknowledges Ministry of Education, Govt. of India, for the Prime Minister’s Research Fellowship, as well as CNRS (Centre national de la recherche scientifique) for postdoctoral research grants.


%

\end{document}